\documentclass[aip,reprint]{revtex4-1}

\usepackage{graphicx}
\usepackage{dcolumn}
\usepackage{bm}
\usepackage{gensymb}
\usepackage[articletitle=true]{achemso}

\draft

\begin{document}

\preprint{AIP/123-QED}

\title{Liquid-Like Interfaces Mediate Structural Phase Transitions in Lead Halide Perovskites}

\author{Connor G. Bischak}
\altaffiliation{These authors contributed equally to this work.}
\affiliation{Department of Chemistry, University of California, Berkeley, CA 94720.}

\author{Minliang Lai}
\altaffiliation{These authors contributed equally to this work.}
\affiliation{Department of Chemistry, University of California, Berkeley, CA 94720.}

\author{Dylan Lu}
\affiliation{Department of Chemistry, University of California, Berkeley, CA 94720.}

\author{Zhaochuan Fan}
\affiliation{Department of Chemistry, University of Utah, Salt Lake City, Utah 84112.}

\author{Philippe David}
\affiliation{Department of Chemistry, University of Utah, Salt Lake City, Utah 84112.}

\author{Dengpan Dong}
\affiliation{Department of Materials Science and Engineering, University of Utah, Salt Lake City, Utah 84112.}

\author{Hong Chen}
\affiliation{Department of Chemistry, University of California, Berkeley, CA 94720.}
\affiliation{School of Environmental Science and Engineering, Southern University of Science and Technology, Shenzhen, Guangdong, China.}

\author{Ahmed S. Etman}
\affiliation{Department of Materials and Environmental Chemistry (MMK), Stockholm University, SE 106 91 Stockholm, Sweden.}
\affiliation{Department of Chemistry, Faculty of Science, Alexandria University, Ibrahimia, 21321 Alexandria, Egypt.}

\author{Teng Lei}
\affiliation{Department of Chemistry, University of Utah, Salt Lake City, Utah 84112.}
\author{Junliang Sun}
\affiliation{College of Chemistry and Molecular Engineering, Peking University, Beijing 100871, China.}

\author{Michael Gr\"{u}nwald}
\affiliation{Department of Chemistry, University of Utah, Salt Lake City, Utah 84112.}

\author{David T. Limmer}
\affiliation{Department of Chemistry, University of California, Berkeley, CA 94720.}
\affiliation{Materials Sciences Division, Lawrence Berkeley National Laboratory, Berkeley, CA 94720.}
\affiliation{Kavli Energy NanoScience Institute, Berkeley, CA 94720.}

\author{Peidong Yang}
\altaffiliation{Corresponding Author (p\textunderscore yang@berkeley.edu)}
\affiliation{Department of Chemistry, University of California, Berkeley, CA 94720.}
\affiliation{Materials Sciences Division, Lawrence Berkeley National Laboratory, Berkeley, CA 94720.}
\affiliation{Kavli Energy NanoScience Institute, Berkeley, CA 94720.}
\affiliation{Department of Materials Science and Engineering, University of California, Berkeley, CA 94720.}

\author{Naomi S. Ginsberg}
\altaffiliation{Corresponding Author (nsginsberg@berkeley.edu)}
\affiliation{Department of Chemistry, University of California, Berkeley, CA 94720.}
\affiliation{Materials Sciences Division, Lawrence Berkeley National Laboratory, Berkeley, CA 94720.}
\affiliation{Department of Physics, University of California, Berkeley, CA 94720.}
\affiliation{Kavli Energy NanoScience Institute, Berkeley, CA 94720.}
\affiliation{Molecular Biophysics and Integrative Bioimaging Division, Lawrence Berkeley National Laboratory, Berkeley, CA 94720.}

\begin{abstract}
Microscopic pathways of structural phase transitions are difficult to probe because they occur over multiple, disparate time and length scales. Using \textit{\textit{in situ}} nanoscale cathodoluminescence microscopy, we visualize the thermally-driven transition to the perovskite phase in hundreds of non-perovskite phase nanowires, resolving the initial nanoscale nucleation and subsequent mesoscale growth and quantifying the activation energy for phase propagation. In combination with molecular dynamics computer simulations, we reveal that the transformation does not follow a simple martensitic mechanism, and proceeds via ion diffusion through a liquid-like interface between the two structures. While cations are disordered in this liquid-like region, the halide ions retain substantial spatial correlations. We find that the anisotropic crystal structure translates to faster nucleation of the perovskite phase at nanowire ends and faster growth along the long nanowire axis. These results represent a significant step towards manipulating structural phases at the nanoscale for designer materials properties.

\end{abstract}


\maketitle

Transformations between different crystal structures shape the properties of solids from the smallest nanocrystals to the interior of planets. Structural phase transitions occur through complex microscopic mechanisms that span several time and length scales. \cite{Hanneman1964High,Tolbert1994Size,Murakami2004Post-Perovskite, Grünwald2006Mechanisms,Khaliullin2011Nucleation,Narayan2016Reconstructing} These microscopic pathways differ markedly from material to material because of the highly anisotropic nature of crystals. At two extremes are martensitic and diffusive transformations. Whereas martensitic transformations involve elastic deformations in each unit cell and can be extremely fast, diffusive transformations occur when the required atomic reorganization is substantial and are therefore typically much slower. Despite the availability of many compatibly slower characterization tools (e.g.\,\textit{in situ} X-ray diffraction (XRD) \cite{Hanneman1964High,Tolbert1994Size,Murakami2004Post-Perovskite} or transmission electron microscopy (TEM)\cite{Zheng2011Observation,Lin2014Atomic}) the mechanistic dynamics of diffusive solid-solid phase transitions have remained elusive. To span the substantial structural mismatch during the transition, the presence of incoherent, disordered interfaces between phases has been hypothesized in a handful of systems—directly observed only in colloidal models,\cite{Peng2015Two-step} simulated in ice-clathrate \cite{Nguyen2015Structure} and tungsten \cite{Barmak2017Transformation} systems, and suggested as mechanistic features in binary metal alloys under the framework of ``massive transformations.''\cite{Howe2002Static,Aaronson2002Mechanisms,Yanar2002Massive} Due to the rugged free energy landscapes of non-trivial atomic rearrangements, the analysis of structural transformations via computer simulations is also challenging because one must incorporate time scales of not only atomistic fluctuations and macroscopic collective reorganizations but also those of processes occurring on scales that are intermediate to these disparate ones, such as diffusion.

Here we reveal the mechanism of a non-martensitic, massive structural transformation on the nanoscale by exploiting differences in cathodoluminescence (CL) emission of different crystal structures and by realizing multiscale simulations that include molecular dynamics (MD) and coarse-grained models. With \textit{\textit{in situ}} scanning electron CL imaging, we follow a thermally induced structural transformation in CsPbIBr\textsubscript{2} nanowires with exquisite temporal and spatial resolution over a large field of view, allowing the simultaneous characterization of the nucleation and growth kinetics in a statistically significant number of single particles. The multi-scale nature of our experimental observations and modeling enables us to determine the microscopic mechanisms of the structural transformation. We find that both nucleation and growth are characterized by strongly anisotropic kinetics and that the boundary between crystal structures is propagated by activated ion diffusion through a liquid-like interface. Despite the overall disorder of the interface, long-range anionic charge density correlations are observed, which promote the crystallographic registration of the two crystal phases. Our observations provide an unprecedented view of the complex dynamic pathways by which two distinct crystal structures interconvert.

\begin{figure*}
\includegraphics[width=14cm]{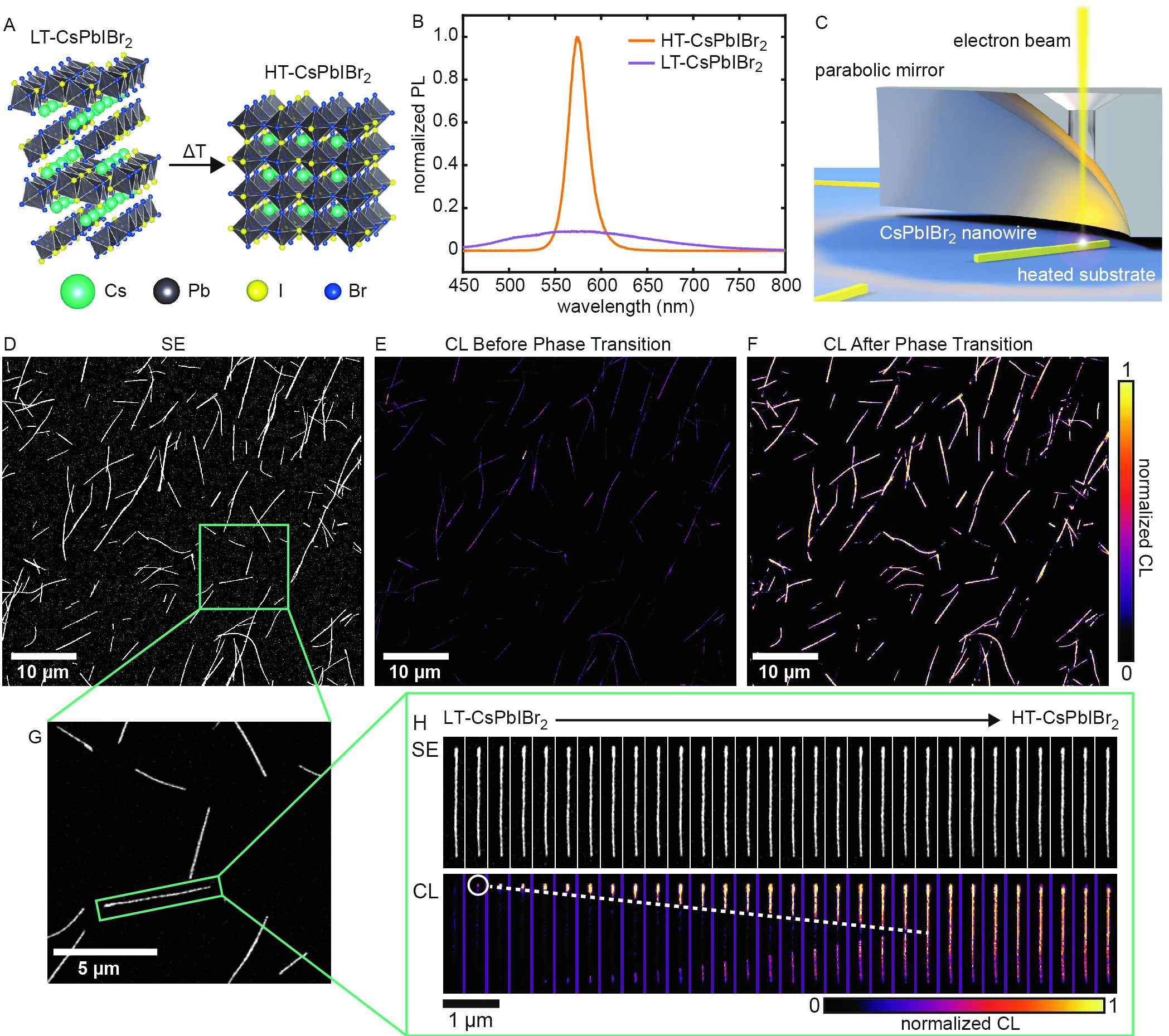}
\caption{\label{fig1}Cathodoluminescence (CL) imaging of the LT-CsPbIBr\textsubscript{2} to HT-CsPbIBr\textsubscript{2} phase transition in nanowires. (A)  Schematic of the LT-CsPbIBr\textsubscript{2} and HT-CsPbIBr\textsubscript{2} phases. (B) Photoluminescence spectra of LT-CsPbIBr\textsubscript{2} and HT-CsPbIBr\textsubscript{2} nanowires. (C) Schematic of the CL imaging apparatus. (D) SE image of the full field-of-view of the measurement containing $\sim$150 individual nanowires. CL images (E) before and (F) after the phase transition. (G) Magnified view showing individual nanowires. (H) An example time series of SE and CL images of a single nanowire during the phase transition at 163 \degree C. The white circle indicates one of two nucleation events in the nanowire, and the dotted white line marks the corresponding phase boundary as it migrates along the length of the nanowire. (Adjacent snapshots are separated in time by 6 seconds.)}
\end{figure*}

Mixed halide perovskites are promising semiconductor materials for optoelectronics\cite{Deschler2014High,Eperon2016Perovskite-perovskite,McMeekin2016mixed-cation} and are an ideal platform for investigating solid-solid phase transitions. Much like CsPbI\textsubscript{3},\cite{Sharma1992Phase,Stoumpos2013Semiconducting, Eperon2015Inorganic,Lai2017Structural} the metal halide perovskite material, CsPbIBr\textsubscript{2}, undergoes a thermally-driven phase transition from a low temperature non-perovskite phase (LT-CsPbIBr\textsubscript{2}) to a high temperature perovskite phase (HT-CsPbIBr\textsubscript{2}) upon heating (Figure 1A) and remains kinetically-trapped in the HT-CsPbIBr\textsubscript{2} phase upon rapid cooling. This phase transition contributes to the instability of perovskite photovoltaics\cite{Eperon2015Inorganic, Yi2016Entropic} and has been utilized for generating nanoscale p-n heterojunctions \cite{Kong2018Phase-transition–induced} and thermochromic smart window technologies. \cite{Lin2018Thermochromic} Despite much interest, the mechanism of this phase transition has not been explained. In contrast to structurally similar perovskite phases known to interconvert,\cite{Dobrovolsky2017Defect-induced, Quarti2016Structural} LT-CsPbIBr\textsubscript{2} and HT-CsPbIBr\textsubscript{2} are not related by simple elastic deformations. Transitions between these structurally dissimilar lattices thus require a more complex rearrangement of atoms, suggesting the possibility to observe a massive transformation and to evaluate how the anharmonic, soft, and ionic nature \cite{Zhu2016Screening} of the material affects its dynamics. Additionally, the density of the material increases by approximately 7\% during the transition, suggesting that disparate interfacial energies and lattice strain could play an important role in nucleation and growth. Crucial to this study, the LT-CsPbIBr\textsubscript{2} phase has an indirect band gap, resulting in low photoluminescence (PL) emission, whereas the direct bandgap HT-CsPbIBr\textsubscript{2} phase exhibits bright PL emission (Figure 1B). This large difference in emission intensity produces strong contrast differences between the two structures during CL imaging and allows us to accurately track the progress of the phase transition.

We use CL imaging with \textit{in situ} heating to monitor the phase transition kinetics of CsPbIBr\textsubscript{2}. CL microscopy uses a focused, scanning electron beam to excite the sample of interest, and the emitted light is collected by a parabolic mirror and directed to a detector to form an image (Figure 1C). It has been used to image steady-state properties \cite{Lai2017Structural,Kong2018Phase-transition–induced,Bischak2015Heterogeneous,Dou2015Atomically,Dar2017Function,Ummadisingu2018Revealing} and dynamic processes\cite{Bischak2017Origin,Bischak2018Tunable} of metal halide perovskites. Similar to with CsPbI\textsubscript{3} nanowires, single-crystal LT-CsPbIBr\textsubscript{2} nanowires were synthesized with the edge-sharing octrahedral chains oriented along the long axis of the nanowires, as confirmed by continuous rotation electron diffraction (cRED) (Figure S1). Using CL microscopy, we measure a large field-of-view containing tens to hundreds of these CsPbIBr\textsubscript{2} nanowires to build up statistics on nucleation and growth kinetics. A secondary electron (SE) image of a typical field-of-view is shown in Figure 1D, containing $\sim$150 nanowires that are monitored simultaneously; CL images showing the nanowires before and after the phase transition are shown in Figure 1E and Figure 1F, respectively. The evolution of the nanowires from the LT-CsPbIBr\textsubscript{2} to the HT-CsPbIBr\textsubscript{2} phase from this larger field-of-view is depicted in Figure S2 and Movie S1. We select individual nanowires from this larger field-of-view (Figure 1G) and measure phase propagation rates. Figure 1H shows a time series of simultaneously acquired SE and CL images that illustrate the evolution from LT-CsPbIBr\textsubscript{2} to HT-CsPbIBr\textsubscript{2} in a single nanowire upon heating. Although no changes are obvious in the SE images, the CL images initially show a nanowire entirely composed of LT-CsPbIBr\textsubscript{2} followed by nucleation of the bright HT-CsPbIBr\textsubscript{2} phase at the nanowire ends and phase propagation along the length of the wire until the wire is completely converted to HT-CsPbIBr\textsubscript{2}. Temperature-dependent in-situ X-ray diffraction (XRD) before and after the phase transition confirms the presence of the two phases (Figure S3).

\begin{figure*}
\includegraphics[width=14cm]{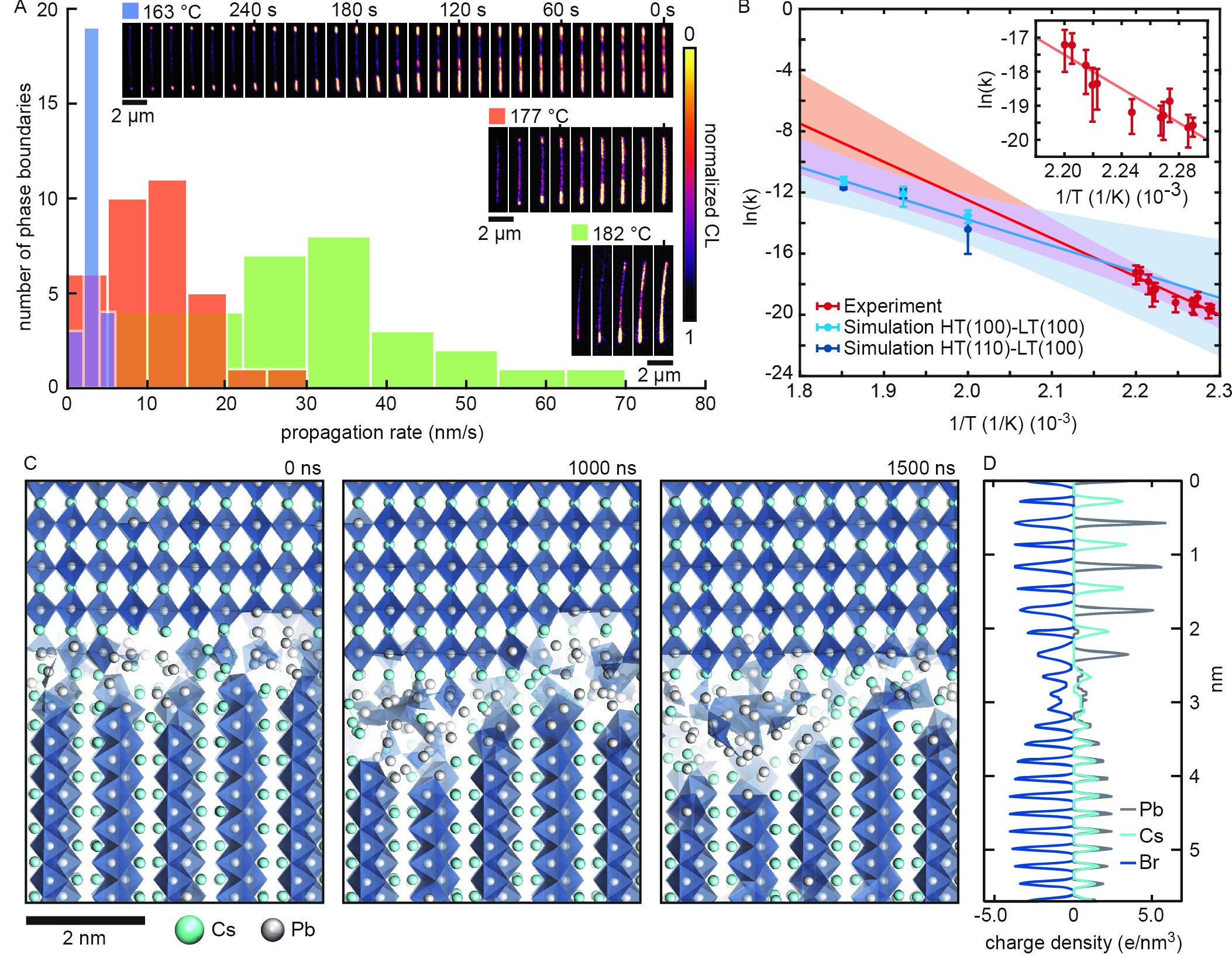}
\caption{\label{fig2} Energetics of perovskite phase propagation. (A) Histograms of the propagation rate of three different populations of nanowires heated at three different constant temperatures, 163 \degree C (purple), 177 \degree C (orange), and 182 \degree C (green). The insets show a characteristic nanowire at each temperature with a countdown time axis for full conversion into the perovskite phase shown at the top. (B) An Arrhenius plot of experimental and MD simulation data. Red points correspond to the experimental propagation rate of different nanowire populations measured at different temperatures, where the units of k are m/s, also shown in the inset; from the simulation data, light blue points indicate growth rates of \textless100\textgreater{}  HT-CsPbBr\textsubscript{3} along the wire axis, as observed in c-RED and SAED (Figure S1), and dark blue points indicate growth rates of \textless110\textgreater{}  HT-CsPbBr\textsubscript{3} along the wire axis. The solid red line is the linear fit to the experimental data (210 $\pm$ 60 kJ/mol), and the solid blue line is the linear fit to the simulation data (140 $\pm$ 80 kJ/mol). The red and blue shaded regions show the 95\% confidence interval of the experimental and simulation fits, respectively, with the overlapping (mauve) region showing the overlap between these two regions. While the \textless110\textgreater{} HT-CsPbBr\textsubscript{3} plane is not observed to be transverse to the wire axis in SAED, the growth rates thus obtained are statistically indistinguishable and further constrain the fit to simulation data. See Figure S16 for further details. (C) Snapshots from the MD simulation of phase propagation as a function of time at 267 \degree C, showing the disordered interface between the LT-CsPbBr\textsubscript{3} and HT-CsPbBr\textsubscript{3} phases. (D) Charge density profiles from the MD simulation at 267 \degree C, obtained by projecting ion positions onto the direction of the wire axis. Data are averaged over a 50 ns time window.}
\end{figure*}

The combination of the nanowire geometry and the high spatial resolution of CL imaging allows for a quantitative analysis of the phase propagation rates along the length of the nanowire. Taking advantage of the ability to probe the phase transition dynamics in many CsPbIBr\textsubscript{2} nanowires simultaneously, we provide a statistical analysis of the HT-CsPbIBr\textsubscript{2} phase growth rate as a function of temperature. We record the phase propagation dynamics of a population of nanowires at temperatures ranging from 163 to 182 \degree C (Figure S4). We first rapidly increase the stage temperature to a specific value to initiate the phase transition and then maintain a constant temperature to measure the phase propagation rate (see Supplementary Information and Figure S5 for more details). Figure 2A shows the distribution of phase propagation rates at three different temperatures with typical time series of individual nanowires shown in the inset. At 163 \degree C, 177 \degree C, and 182 \degree C, we observe average propagation rates of 3.1 $\pm$ 0.2 nm/s, 11 $\pm$ 1 nm/s, and 33 $\pm$ 3 nm/s, respectively. This strong temperature dependence suggests that phase propagation is controlled by thermally activated microscopic processes. Indeed, we find that the propagation rate as a function of temperature exhibits Arrhenius-like behavior with an activation energy of 210 $\pm$ 60 kJ/mol (Figure 2B)—~50$\times$ the scale of a typical thermal fluctuation ($k_{\rm B} T$). (Uncertainty provided represents 95\% confidence interval.)  Given this substantial energetic barrier, there must be a significant compensating increase in entropy for interphase boundary propagation to proceed at the observed rates.

\begin{figure*}
\includegraphics[width=14cm]{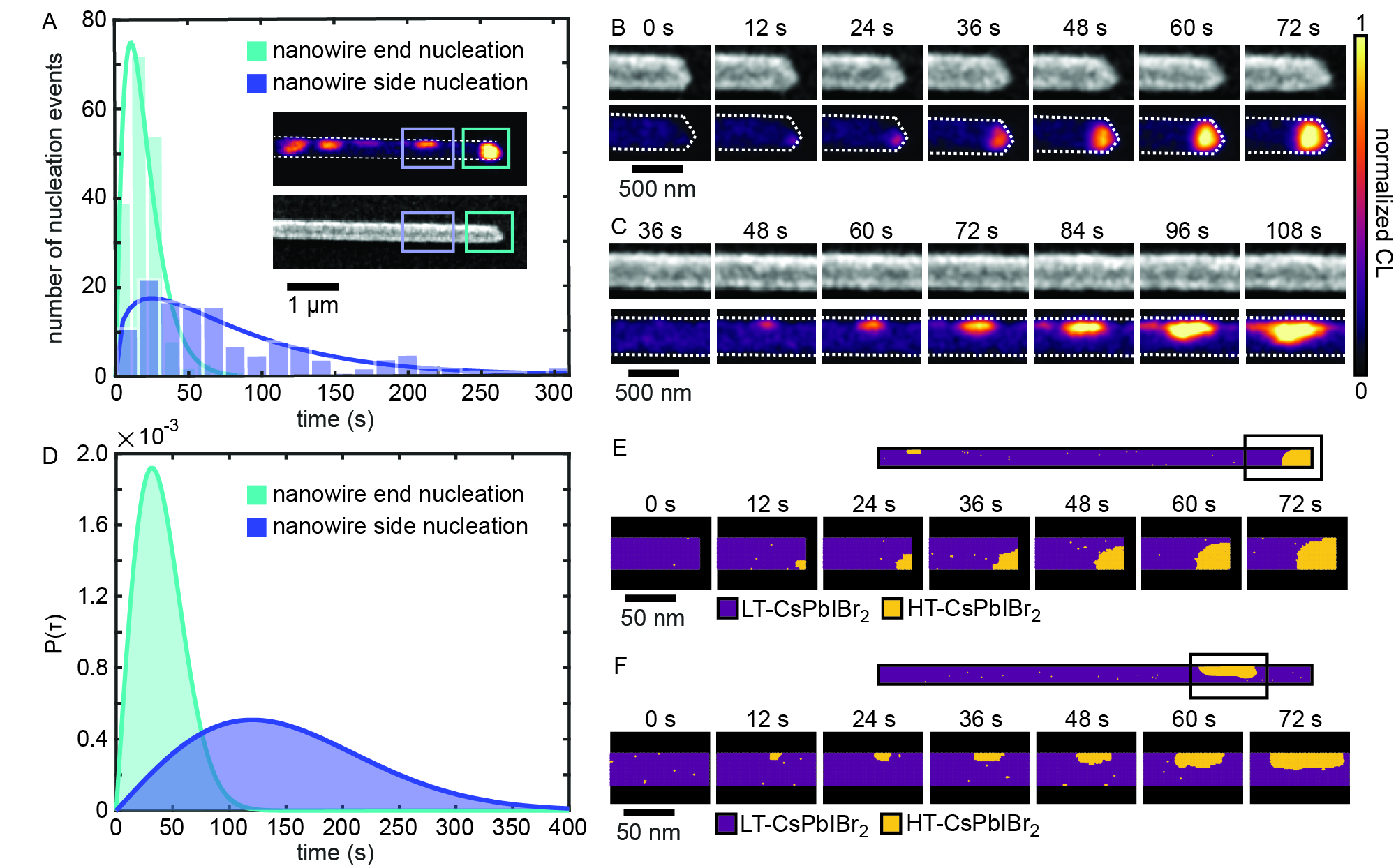}
\caption{\label{fig3} Perovskite phase nucleation and anisotropic growth from experiment and simulation. (A) Plot of observed nucleation events as a function of time for nucleation at nanowire ends (teal), and nucleation at nanowire sides (blue). (B) A time series of CL and SE images of a single nucleation event at the end of a nanowire. (C) A time series of CL and SE images of a single nucleation event at the side of a nanowire. An outline of nanowire from the corresponding SE image is shown in the CL images in B and C. (D) Plot of nucleation probability (P($\tau$)) as a function of time for nucleation at nanowire ends (teal) and nanowire sides (blue) as a function of time from simulations. (E) A time series showing a single nucleation event at the nanowire end from simulations. (F) A time series showing a single nucleation event at the nanowire side from simulations.}
\end{figure*}

To reveal the microscopic mechanism of phase propagation, we use electronic structure calculations to parameterize a classical force field for CsPbBr\textsubscript{3} crystals, which we use as a proxy for the structurally similar mixed halide perovskite studied experimentally (see Table S1 and Table S2, Figure S6 and Figure S7, and the Supplementary Information). We employ this force field in MD simulations. We start our simulations from configurations that include planar interfaces between the LT-CsPbBr\textsubscript{3} and HT-CsPbBr\textsubscript{3} phases, employing the relative crystallographic orientations suggested by single area electron diffraction (SAED) ((100) LT-CsPbBr\textsubscript{3} abutting (100) HT-CsPbBr\textsubscript{3}), as seen in Figure 2C. We considered additional interfacial possibilities, such as (110) LT-CsPbBr\textsubscript{3} planes abutting (100) HT-CsPbBr\textsubscript{3}, described further in the Supplementary Information. In several long MD simulations at temperatures of 227 \degree C, 247 \degree C, and 267 \degree C we consistently observe growth of the HT-CsPbBr\textsubscript{3} phase, in agreement with experiments (see Figures S8-S10 and Supplementary Information). (Phase propagation was intractably slow in simulations at lower temperatures.) Propagation rates estimated from at least three independent simulations for each aforementioned interface at each temperature are plotted in Figure 2B. An Arrhenius-type analysis of these data yields an activation energy of 140 $\pm$ 80 kJ/mol, in good agreement with experiments.

The simulations reveal the formation of a structurally disordered, liquid-like interfacial layer as the origin of the compensating increase in entropy hypothesized from the experiments as necessary to achieve the observed interphase boundary propagation rates. Large energetic barriers to phase propagation, as measured in our experiments and simulations, can be interpreted as describing activated ion diffusion events between stoichiometrically different coordination environments. Growth of the HT-CsPbBr\textsubscript{3} structure proceeds via diffusion of ions across this amorphous interface (Figure 2C and Movie S2).\cite{Peng2015Two-step} Due to the marked structural differences of the two phases, growth cannot proceed layer-by-layer. In particular, (100) layers of HT-CsPbBr\textsubscript{3} have a different ionic composition than (100) layers of LT-CsPbBr\textsubscript{3}. Completion of a new layer of HT-CsPbBr\textsubscript{3} thus requires the recruitment of ions from at least two (100) layers of LT-CsPbBr\textsubscript{3}.

Despite the mediating liquid-like region between the two phases, using SAED we find the same crystallographic orientation of the HT phase in all nanowires probed. This can be understood by separately computing the distribution of each species of ion within the disordered region in the molecular simulations (Figure 2D). Interestingly, we find that, while the positions of the cations are essentially uniform, the positions of the halides remain correlated across the entire liquid-like interface. These correlations manifest as oscillations in the halide density distribution, shown in Figure 2D, and act to template the formation of the growing HT phase, ensuring its crystallographic orientation despite the overall incoherent interface. These persistent anionic density correlations are reminiscent of those templating the  association of metal nanoparticles in dense ionic solutions.\cite{Zhang2017Stable}

In our experiments, we observe nucleation of the HT-CsPbIBr\textsubscript{2} primarily at nanowire ends; a minority of wires also displays nucleation along the lateral surface. Preferential nucleation on nanowire ends is evident from Figure 3A, which shows a histogram of all nucleation events in ~100 nanowires as a function of time, observed at 163 \degree C (see Supplementary Information and Figure S11 for more details). Here, time zero is defined by the image frame in which we first observe a single HT-CsPbIBr\textsubscript{2} phase pixel above the LT-CsPbIBr\textsubscript{2} background CL intensity threshold. The distributions of end and side nucleation events are clearly different; the earlier peak of the end nucleation distribution indicates faster nucleation at nanowire ends than sides. To extract a nucleation rate at each of the nanowire ends and sides, we examine distributions of waiting times, which are exponentially distributed, as expected for an independent, reaction limited process. We find that nucleation at nanowire ends occurs with a rate of 0.045 s\textsuperscript{-1}, whereas nucleation at wire sides occurs with a rate of 0.013 s\textsuperscript{-1} (Figure S12).  The nucleation rates at early times are consistent with those inferred at latter times using an Avrami analysis,\cite{Chaikin2000Principles} accounting for a constant rate of growth measured independently (see Supplementary Information for details). The inset of Figure 3A shows CL and SE images of part of a single nanowire after nucleation occurred at the end and side of the nanowire. Time series of the CL and SE images of the nuclei growth at the nanowire end and nanowire side are depicted in Figure 3B and 3C, respectively (Movie S3 and S4). Additional nucleation events are shown in Figure S13. Nucleation at the nanowire end leads to the new phase expanding until it occupies the full width of the nanowire and then propagating along the long axis of the wire (Figure 3B). Because axial growth can occur in two opposite directions, new regions of HT-CsPbIBr\textsubscript{2} formed on the nanowire side are seen to clearly grow asymmetrically with a faster growth rate along the long axis of the nanowire (Figure 3C). By analyzing the extent to which the perovskite phase is circular in the image frames at early times, we find that nucleation at the nanowire sides results in a more anisotropically-shaped perovskite phase volume with faster growth along the length of the wire (Figure S14).

We develop a phenomenological model to describe both the preferential location of nucleation sites at the wire end as well as the anisotropic growth rate at the nanowire side (see Figure S15 and Supplementary Information for additional details). Specifically, we use a simple lattice model of crystal growth with anisotropic bond and surface energies that are greater along the long axes of the wire compared to the short axes. This model phenomenologically accounts for the anisotropy in lattice orientations of the HT-CsPbIBr\textsubscript{2} and LT-CsPbIBr\textsubscript{2} phases. The lattice is initialized in the pure LT-CsPbIBr\textsubscript{2} phase and is instantly quenched to favor the HT-CsPbIBr\textsubscript{2} phase. Figure 3D shows two distributions of nucleation events at nanowire ends and nanowire sides from the model, which qualitatively matches our experimental results, suggesting that preferential nucleation at nanowire ends is caused by the tendency to minimize interfacial energies between the two phases. Two time series of simulation snapshots of end and side nucleation are shown in Figure 3E and 3F. Growth from the nanowire end proceeds almost isotropically, whereas growth on the nanowire side occurs anisotropically due to the alternation between octahedral double chains and gaps in between them. Specifically, growth occurs at a faster rate in the direction of the lead-halide octahedral chains (i.e., along the wire axis); growth in perpendicular directions is noticeably slower. Our MD simulations confirm these anisotropic growth rates: Interfaces involving (010) and (001) layers of LT-CsPbBr\textsubscript{3}, which are parallel to the nanowire axis, display slower average phase propagation compared to orthogonal (100) layers (Figure S6).

The kinetic pathways of the structural phase transition are also apparent in the morphology and photophysical properties of the resulting HT-CsPbIBr\textsubscript{2} nanowires. In some cases, when the nanowire is not uniformly contacting the substrate, changes in the nanowire morphology are evident upon the formation of the perovskite phase. As the phase transition proceeds, lattice stress associated with the 7\% increase of nanowire volume results in nanowire bending and expansion (Figure S7). Additionally, HT-CsPbIBr\textsubscript{2} nanowires with multiple nucleation sites sometimes display dark regions where two phase boundaries meet (Figure S8), suggesting the formation of grain boundaries with suppressed emission due to a high defect concentration.

Through \textit{in situ} dynamical CL imaging and multiscale modeling of the LT-CsPbIBr\textsubscript{2} to HT-CsPbIBr\textsubscript{2} structural phase transition we have uncovered the mechanism for a complex, non-martensitic structural transformation. The measured activation energy for phase propagation is consistent with a disordered interface between the two phases through which ions must diffuse, as observed in MD simulations. This liquid-like interface is far from the melting point of the involved solids and presents strong anion density correlations, which we suspect as being responsible for the fixed crystallographic orientation of the nascent perovskite phase within the nanowire that is observed in SAED. The spontaneous formation of an incoherent interface suggests that an ordered solid-solid interface between these two structurally dissimilar phases is thermodynamically less favorable than the sum of LT-CsPbIBr\textsubscript{2}-liquid and HT-CsPbIBr\textsubscript{2}-liquid interfaces, plus the concomitant enthalpy required to disorder the interfacial layer, even when such liquid-like configurations are not stable by themselves. Our findings are yet another manifestation of the liquid-like dynamics that have been observed in these highly anharmonic metal halide perovskite lattices that result from the inherent low cohesive energy of their ionic bonds, in contrast with traditional behaviors of covalent semiconductors.

Our results suggest that similar diffusive mechanisms might occur in a large range of materials, including other perovskite materials, that crystallize in structures that do not share simple epitaxial interfaces. The experimental method for observing dynamic structural changes introduced in this work could also be extended to other systems, such as 2D transition metal dichalcogenides\cite{Tongay2012Thermally,Bediako2018Heterointerface} and metal organic frameworks (MOFs),\cite{Silva2010Metal–organic,Cui2012Luminescent} in which a change in the luminescence intensity or wavelength accompanies a structural change. We expect that similar \textit{in situ} monitoring of phase transitions will significantly aid our ability to characterize phase behavior, enabling quantitative comparison to theoretical results and creating opportunities to manipulate solids and their properties on the nanoscale.

\section*{Acknowledgments}

We thank E. Wong, E. S. Barnard, D. F. Ogletree, and S. Aloni at the Molecular Foundry for assistance with CL equipment and helpful discussions. CL and analysis work by C.G.B. and N.S.G. has been supported by STROBE, A National Science Foundation Science \& Technology Center under Grant No. DMR 1548924. The CL imaging at the Lawrence Berkeley Lab Molecular Foundry and the TEM imaging at the National Center for Electron Microscopy were performed as part of the Molecular Foundry user program, supported by the Office of Science, Office of Basic Energy Sciences, of the U.S. Department of Energy under Contract No. DE-AC02-05CH11231. C.G.B. acknowledges an NSF Graduate Research Fellowship (No. DGE1106400), and N.S.G. acknowledges an Alfred P. Sloan Research Fellowship, a David and Lucile Packard Foundation Fellowship for Science and Engineering, and a Camille and Henry Dreyfus Teacher-Scholar Award. Modeling by D.T.L., materials fabrication and characterization by P.Y. and co-workers is supported under the U.S. Department of Energy, Office of Science, Office of Basic Energy Sciences, Materials Sciences and Engineering Division under Contract No. DE-AC02-05-CH11231 within the Physical Chemistry of Inorganic Nanostructures Program (KC3103). Molecular dynamics simulations have also been partially supported by the National Science Foundation under NSF-REU grant CHE-1659579. The support and resources of the Center for High Performance Computing at the University of Utah are gratefully acknowledged.

\section*{Author Contributions}

C.G.B, M.L., D.T.L., M.G., P.Y., and N.S.G. wrote the manuscript. C.G.B., M.L., D.Y., D.T.L., P.Y., and N.S.G. conceptualized the experiment. M.L. synthesized the LT-CsPbIBr\textsubscript{2} nanowires and performed the PL and XRD measurements. C.G.B. performed the CL experiments and analyzed the CL nucleation and growth data. D.T.L. conceptualized and performed the Ising model simulations. Z.F., P.D., D.T.L., and M.G. conceived the MD simulations and analyzed simulation results. Z.F., P.D., D.D., and M.G. developed the force field and performed MD simulations. A.S.E., J.S., H.C. collecting and analyzing the continuous rotation electron diffraction (cRED). T.L. performed the SAED.

\section*{References}



\end{document}